\documentclass{article}
\usepackage{graphicx} 
\usepackage{authblk}
\usepackage{amsmath}
\usepackage{multirow}
\usepackage[sort, comma, numbers]{natbib}
\usepackage{pdflscape}
\usepackage[skip=5pt plus1pt, indent=40pt]{parskip}

\usepackage{geometry}
\usepackage{fancyhdr}
\title{Quartered Spectral Envelope and 1D-CNN-based Classification of Normally Phonated and Whispered Speech}
\author[1]{S. Johanan Joysingh}
\author[1]{P. Vijayalakshmi}
\author[2]{T. Nagarajan}
\affil[1]{Department of ECE, Sri Sivasubramaniya Nadar College of Engineering, Chennai, India.}
\affil[2]{Department of CSE, Shiv Nadar University Chennai, India.}
\date{}

\begin{document}
\maketitle

\section{Abstract}
Human-computer interaction via speech is more common than ever before.
Whisper, as a form of speech, is not sufficiently addressed by mainstream speech applications, such as automatic speech recognition, speaker identification, language identification, etc, even though there are more than a hundred thousand laryngectomees in the world who can only whisper.
This is due to the fact that systems built for normal speech do not work as expected for whispered speech.
A first step to building a speech application that is inclusive of whispered speech, is the successful classification of whispered speech and normal speech.
Such a front-end classification system is expected to have high accuracy and low computational overhead, which is the scope of this paper. 
One of the characteristics of whispered speech is the absence of the fundamental frequency (or pitch), and hence the pitch harmonics as well. 
The presence of the pitch and pitch harmonics in normal speech, and its absence in whispered speech, is evident in the spectral envelope of the Fourier transform.
We observe that this characteristic is predominant in the first quarter of the spectrum, and exploit the same as a feature. 
We propose the use of one dimensional convolutional neural networks (1D-CNN) to capture these features from the quartered spectral envelope (QSE).
The system yields an accuracy of 99.31\% when trained and tested on the wTIMIT dataset, and 100\% on the CHAINS dataset.
The proposed feature is compared with Mel frequency cepstral coefficients (MFCC), a staple in the speech domain. 
The proposed classification system is also compared with the state-of-the-art system based on log-filterbank energy (LFBE) features trained on long short-term memory (LSTM) network.
The proposed system based on 1D-CNN performs better than, or as good as, the state-of-the-art across multiple experiments. 
It also converges sooner, with lesser computational overhead.
Finally, the proposed system is evaluated under the presence of white noise at various signal-to-noise ratios and found to be robust.

\vspace{0.25cm}
\begin{center}
\textbf{Keywords}: \textit{whispered speech, pitch harmonics, magnitude spectrum, classification, CNN}
\end{center}

\section{Introduction }
\label{introduction}
Human-computer communication through speech must be as accessible as the interaction between two humans.
In this regard, addressing whispered speech as a form of speech communication between humans and computers is vital.
Whispered speech is a form of speech that does not possess the glottal activity that is commonly associated with normally phonated speech \citep{wilson1985comparative}. 
In other words, it is not voiced. 
This leads to an absence of the fundamental frequency (F0). 
From the viewpoint of the source-system model \cite{rabiner2011theory, quatieri2002discrete, makhoul1975linear} of speech production, normally phonated speech (henceforth called normal speech) is produced by two sources of excitation depending on the phone, 
(i) due to the vibrations produced by the glottis and 
(ii) due to air passing through an open glottis. 
For whispered speech, only the latter form of excitation exists.  

Whispered speech, as a sub-domain of speech technology is important because there are occasions when users can only, or may prefer to, whisper. 
For people who have had a laryngectomy, or people who have issues with their larynx in general, whispering is the only option. 
In a report concluded in 2013, there were about 50,000 laryngectomees in the US alone \citep{brook2013laryngectomee}.
Extrapolating this over time and geography, we can expect this number to be well over hundred thousand in 2022. 
Hence, addressing whispered speech would lead to a more inclusive speech application.
Other applications include taking voice-typed notes from quiet settings such as libraries, hospitals and airplanes, or speech applications used by detectives and army personnel who are on stealth missions, etc.
The importance of whispered speech can also be understood from the fact that Amazon's Alexa addresses whispered speech \citep{cotescu2019voice}.

When a speech technology application handles multiple forms of speech at its input, it may require a front-end classification system that determines the form of speech. 
For example, a language/dialect identification system that acts as a front-end to an automatic speech recognition (ASR) system, to detect the language, so that the corresponding acoustic and language models can be loaded.
The focus of the current work, that is the classification of whispered speech and normal speech, is also a front-end task.
It has also been established that, when models trained on normal speech are used on whispered speech, there is a degradation in performance \citep{ito2005analysis, grozdic2017whispered, jin2007whispering}, hence it is essential to classify these two forms of speech before hand.

There exist some established characteristic differences between whispered speech and normal speech. 
A very comprehensive analysis can be found in \cite{wilson1985comparative} and \cite{lim2011computational}. 
\cite{jovivcic1998formant} and \cite{wenndt2002study} also focus on the analysis of whispered and normal speech.
To summarize their findings here, these differences are mostly based on the location, amplitude, and bandwidth of the formants, as estimated by the linear prediction (LP) spectrum. 
For example, in whispered speech,  
\begin{itemize}
    \item the locations of the formants are shifted higher,
    \item the amplitudes of the formants are lower, and
    \item the bandwidths of the formants are wider.
\end{itemize}

These characteristics were initially utilized in literature using hand-crafted features, in a signal processing framework as, for example, in \cite{wenndt2002study}. 
One drawback with this method is that they required a threshold to work. 
The threshold is estimated based on analysis of whispered and normal speech commonly encountered by the system. 
The other drawbacks are that, 
(i) the locations of the formants varied from person to person, and
(ii) the amplitudes had to be normalized, before a comparison could be made.
More information about this can be found in \cite{wilson1985comparative}.
In light of this, more objective features and/or techniques for classification are preferred. 

Existing work on the classification of whisper and normal speech mainly differs in terms of the feature used for classification, the classifier used for modelling, and the way in which either frame level or utterance level classification is performed. 
The datasets most used are wTIMIT \citep{khoria2021teager, shah2021exploiting, ashihara2019neural} and CHAINS \citep{khoria2021teager, shah2021exploiting}.
In some cases, authors also provide classification results on a corpus created in-house, along with the results for publicly available datasets \citep{shah2021exploiting, raeesy2018lstm}. 
Classifier performance in the presence of noise is also reported in some cases, using a corpus such as MUSAN \citep{shah2021exploiting}.
A series of implementations that involves Gaussian mixture models (GMM) to model a set of hand-crafted features, can be found in \cite{zhang2007analysis, zhang2008entropy, zhang2009advancements}. 
In \cite{zhang2010whisper}, a whisper island detection system based on Bayesian information criterion is proposed.
Authors of \cite{raeesy2018lstm} propose that since deep neural networks (DNN) work well for voice activity detection (VAD), it must work well for whisper speech detection as well.
They propose a whisper speech detection system based on long short-term memory (LSTM). 
There have been a few implementations involving the 2D-CNN, for example \cite{ashihara2019neural, shah2021exploiting} for the current task.
In \citep{baghel2019shouted}, 1D-CNN is used in the classification of normal speech and shouted speech (which is the opposite of whispered speech) using the magnitude spectrum.
To our knowledge, 1D-CNN has not been exploited for the current task.
The use of hand-crafted features and other features like MFCC \citep{ashihara2019neural}, log filter-bank cepstral coefficients (LFCC), Teager energy cepstral coefficients (TECC) \citep{khoria2021teager}, log filter-bank energy (LFBE) \citep{raeesy2018lstm}, and group delay spectrum \citep{shah2021exploiting} are available in literature. 
The maximum accuracy reported for the wTIMIT dataset is 100\% by \cite{shah2021exploiting}, following which \cite{khoria2021teager} report an accuracy of 92.22\%.
In general, what we notice from our literature review is that a lot of focus has been on the formant structure of normal and whispered speech, but little or no attention has been given to the presence of F0 and pitch harmonics.
The F0 and pitch harmonics are more concentrated on the first quarter of the spectrum.
To our knowledge there hasn't been a system based on exploiting the quartered spectral envelope. 
The intuition and motivation behind the proposal, and the technical details, are provided in the following section.

\section{The Proposed System}
\subsection{Analysis} 

\begin{figure}[h]%
\centering
\includegraphics[width=\textwidth]{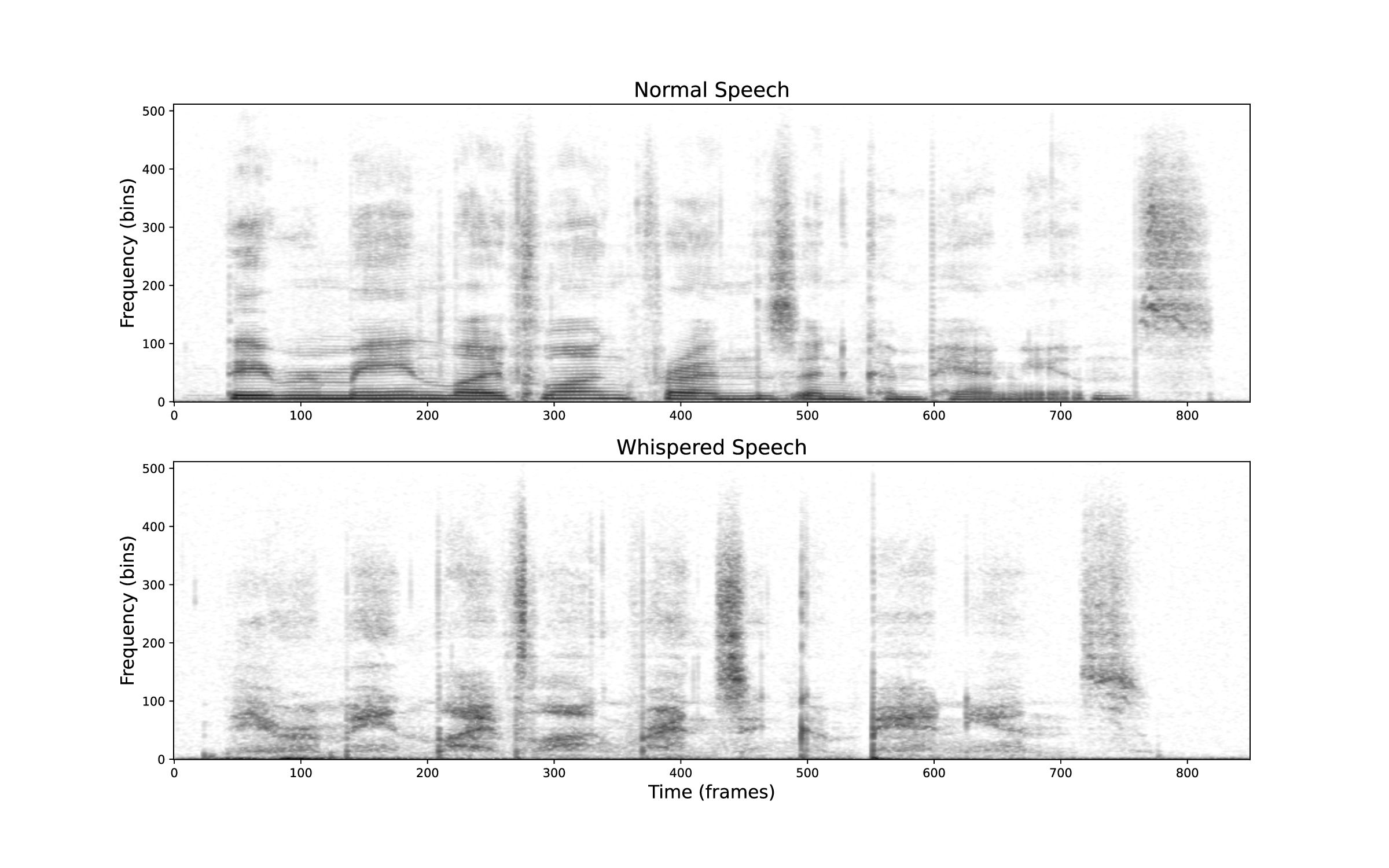}
\caption{Spectrogram computed using a 1024 point FFT, showing the differences between normal (above) and whispered speech (below), sampled at 44.1kHz.}
\label{fig:spectrogram_full}
\end{figure}

Figure \ref{fig:spectrogram_full} shows the spectrogram of normal speech and whispered speech for an utterance corresponding to the sentence --- ``They remain lifelong friends and companions". 
It can be seen from the figure that, due to the presence of the fundamental frequency (or pitch) in normal speech, there are well defined pitch harmonics in the spectrum. 
These are seen as the horizontal striations in the spectrogram. 
While the spectrogram of whispered speech, with the same spectral and temporal resolution, has no such feature due to the absence of pitch. 
The presence of F0 (or tone) in normal speech, and its absence in whispered speech, is established \citep{qian2020tagging}.
Besides the pitch harmonics, differences in formant structure can also be observed in the figure. 
A suppressed formant structure, with increased formant bandwidth, can be observed for whispered speech, while for normal speech, the formants are more pronounced. 

\begin{figure}[h]%
\centering
\includegraphics[width=\textwidth]{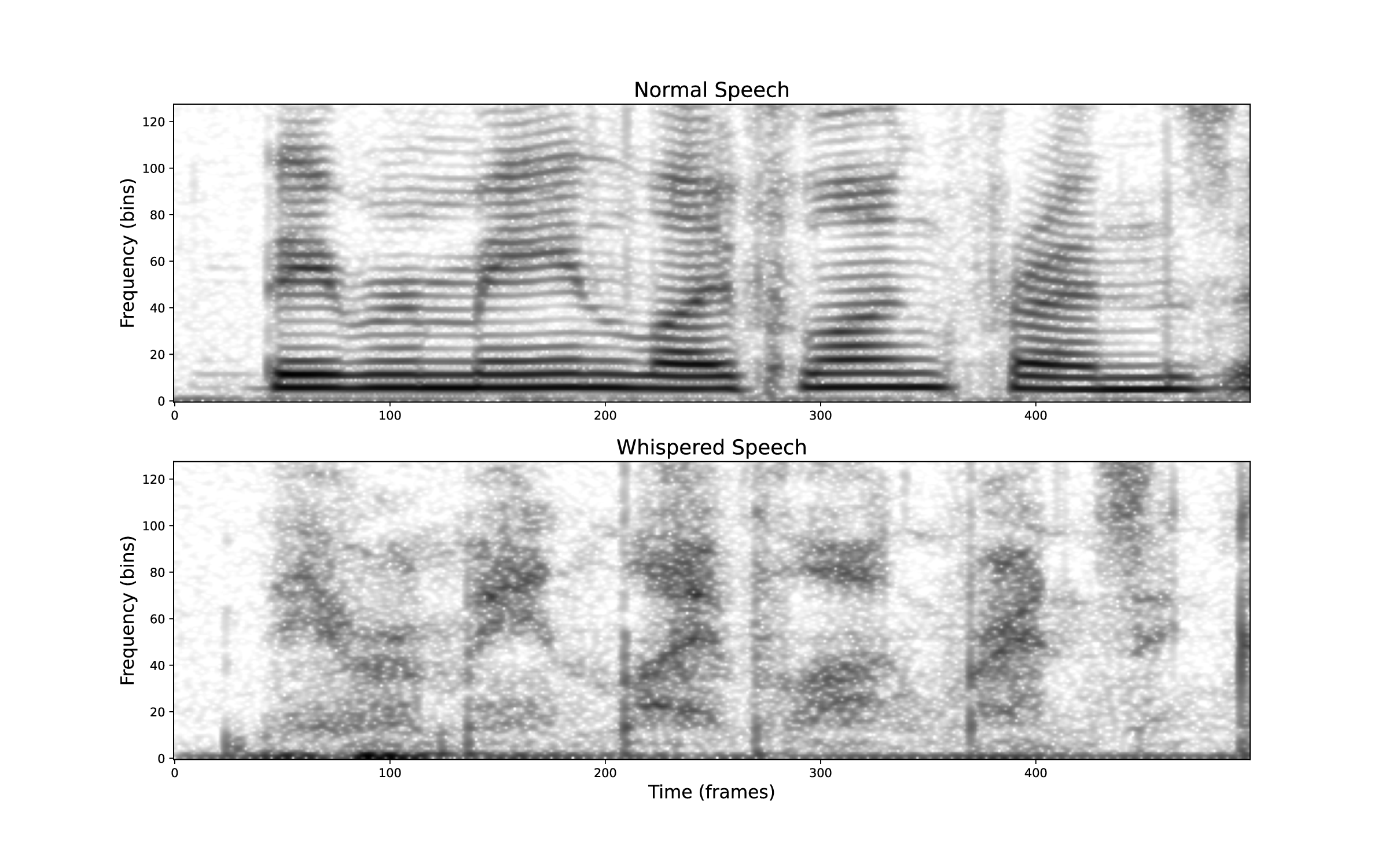}
\caption{Quartered Spectrogram computed using a 1024 point FFT, showing bins ranging from 1 to 128, for normal (above) and whispered speech (below), sampled at 44.1kHz.}
\label{fig:spectrogram_section}
\end{figure}

It, can also be noticed from Fig. \ref{fig:spectrogram_full} that the first quarter of the spectrum contain the more prominent pitch harmonics.
Figure \ref{fig:spectrogram_section}, is a zoomed version of Fig. \ref{fig:spectrogram_full}.
The zoomed version is obtained by slicing the time frames from 80 to 100 in Fig. \ref{fig:spectrogram_full} and, the bins from 1 to 128 (the first quarter) --- a quartered spectrogram.
One slice of this quartered spectrogram, across frequency, is what we call, a quartered spectral envelope (QSE) feature.
The prominence of the pitch harmonics in the first quarter of the spectral envelope in normal speech is evident from this figure.
This is the premise of the proposed feature.

\subsection{Proposal}
\label{section:proposal}
We propose firstly that, the strength and prominence of the pitch harmonics, as expressed by the spectral envelope, can be used as a feature to classify normal speech from whispered speech. 
Secondly that, this feature can be extracted and trained using the one-dimensional convolutional neural network (1D-CNN), which to our knowledge has not been employed for this task. 
Here, the one-dimensional convolution is performed across frequency. 
The hypothesis is that, with the right kernel size, the 1D-CNN can be expected to learn the presence and prominence of the pitch harmonics.
Thirdly that, most of the discriminative information is concentrated in the first quarter of the spectrum, from $0$ to $K/4$ bins, where $2K$ is the size of the fast Fourier transform (FFT). 
Hence, we propose discarding the information corresponding to the higher frequencies since their contribution to the current classification task is insignificant (in the 1D-CNN framework).
The quartered spectrogram is denoted by,
\begin{equation}
    X(n,k) = S(n,k) \qquad 1<k<K/4
\end{equation}
where, $S(n,k)$ is the spectrogram.
Each slice of $X(n,k)$ (for $n = 1,2,\dots,N$), is the QSE, and is the input to the 1D-CNN.
Reducing the dimensions of the input spectra, can be viewed as a feature selection process and it offers two benefits. 
Firstly, it reduces the amount of data required for training to converge.
This is especially important for whispered speech, as the quantity of data available is limited and, collecting whispered speech data is not as trivial as collecting normal speech data. \citep{ashihara2019neural}
Secondly, it is computationally more efficient.

\subsection{1D-CNN Architecture} 
\label{section:1d-cnn}
\begin{landscape}
\begin{figure}
    \centering
    \includegraphics[width=1.3\textwidth]{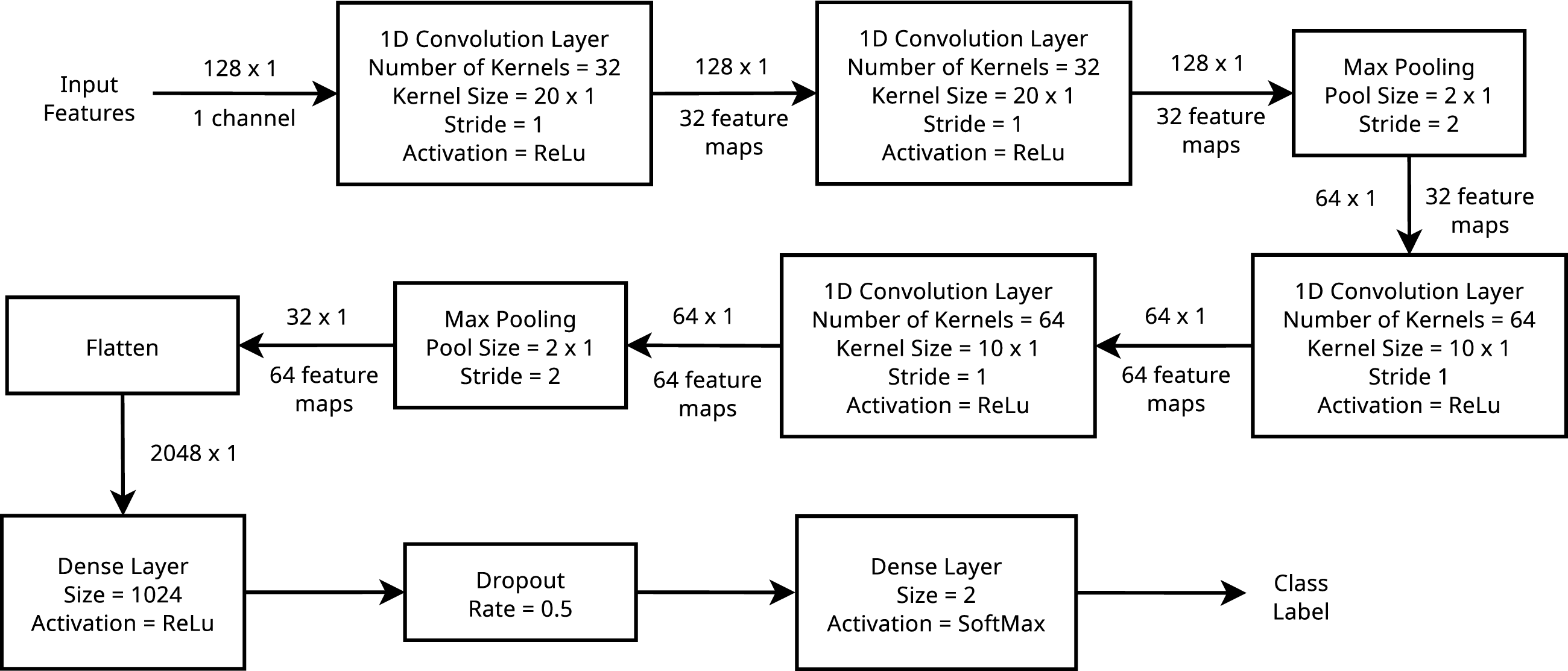}
    \caption{The architecture of the 1D-CNN proposed in the current work. Figure shows the architecture that offered the best results.}
    \label{fig:cnn_arch}
\end{figure}
\end{landscape}

The basic architecture of the 1D-CNN is as shown in Fig. \ref{fig:cnn_arch} and explained below. 
It essentially consists of two sets of 1D convolutional layers, one fully connected layer and one output layer.
Each set of 1D convolutional layers, consist of two convolutional layers and one max-pooling layer. 
The first set of two convolutional layers consists of 32 filters each, while the second set of two convolutional layers consists of 64 filters each.
The kernel size is fixed based on the length (or span) of a pitch harmonic, and can directly affect the performance of the classifier. 
Meaning that the kernel should be large enough to be able to cover the peaks corresponding to a few pitch harmonics, so as to learn their shapes, but not too large to smoothen it out.
Best results were obtained using kernels of size 20x1 and 10x1, for the two sets of convolutional layers respectively. 
In all convolutional layers, zero padding is assumed, such that the size of the input at each layer is maintained. 
The set of convolutional layers are further connected to a fully connected dense layer that actually learns the features extracted by the convolutional layers. 
All convolutional layers, and the first dense layer, use a rectified linear unit (ReLU) activation function.
This is followed by a dropout layer, leading to the dense layer that provides two outputs.
This final dense layer is activated by a softmax function that provides posteriors in the range [0,1].

Over the course of our experiments, we varied the architecture of the 1D-CNN by changing the number of convolutional layers, kernel sizes, and the number of dense layers.
The architecture yielding the best performance is presented in Fig. \ref{fig:cnn_arch}.
More details can be found in Section. \ref{section:training}.

\section{Experimental Setup} 
\label{section:experimental_setup}

The following sections detail the datasets, features, and architectures used for evaluation. 

\subsection{Datasets}
\label{section:datasets}
For evaluation, we use the wTIMIT \citep{lim2011computational} and CHAINS \citep{cummins2006chains} datasets. 
The wTIMIT dataset consists of normal and whispered utterances recorded in two phases. 
The first phase was carried out in Singapore, and the second phase was carried out in the United States (US). 
In the current work, we consider the data collected in the US, as in \cite{shah2021exploiting}.
It is a very large dataset (compared to the CHAINS dataset), and is employed for data intensive tasks such as speech conversion (whisper to normal, vice-versa) in literature \cite{cotescu2019voice}.
The recordings were obtained from 28 speakers from the US, of which 16 were female, and 12 were male.
The CHAINS (characterizing individual speakers) dataset, is originally intended for speech recognition, but since it has a parallel set of normal and whispered utterances, it is also used for the current task \citep{khoria2021teager, shah2021exploiting}. 
The recordings were obtained from 36 speakers, of which 16 were female, and 20 were male.
The utterances in both datasets are recorded in a clean environment, with the sampling rate set to 44.1kHz.
The details of the datasets, with respect to the duration and the number of utterances, are provided in Table \ref{table:dataset}. 
Each utterance has an average duration of 5.20s that consists of a brief moment of silence before and after the utterance. 

\begin{table}[h]
\begin{center}
\caption{Details of the speech datasets used in the current work}
\label{table:dataset}
\vspace{0.25cm}
\begin{tabular}{clcccc}
\hline
\multicolumn{1}{l}{}    &                                     & \multicolumn{2}{c}{Train}             & \multicolumn{2}{c}{Test} \\ \cline{3-6}
\multicolumn{1}{l}{}    &                                     & \multicolumn{1}{l}{Normal}  & Whisper & Normal     & Whisper     \\ \hline
\multirow{2}{*}{wTIMIT} & \multicolumn{1}{c}{Num. Utterances} & 10,934                      & 10,245  & 725        & 723         \\
                        & Duration                            & \multicolumn{1}{l}{15h 31m} & 16h 25m & 1h 19m     & 1h 28m      \\ \hline
\multirow{2}{*}{CHAINS} & \multicolumn{1}{c}{Num. Utterances} & 932                         & 932     & 400        & 400         \\
                        & Duration                            & 1h 46m                      & 1h 48m  & 13m        & 13m        \\ \hline
\end{tabular}
\end{center}
\end{table}

\subsection{Features}
To extract the proposed intermediate features as input to the 1D-CNN, short-time Fourier transform (STFT) is computed with the frame-size set to 1024 and frame-shift set to one eighth of the frame-size.
This comes to 23 and 2ms respectively when the sampling rate is set to 44.1kHz, and 64 and 8ms respectively when the sampling rate is set to 16kHz.
The first quarter of the number of frequency bins, that is up to 128, are retained as features for training the 1D-CNN \cite{nagarajan2004subband, vijayalakshmi2007acoustic, vijayalakshmi2005analysis}, as shown in Section. \ref{section:proposal}.
With the frame-size and frame-shift set to the same values, MFCC features are extracted with the number of Mel bands set to 64. 
The proposed 1D-CNN architecture is trained with these MFCC features.
This gives an idea about how the proposed QSE feature performs with respect to an ubiquitous feature in literature (MFCC).
For comparison with the state-of-the-art system, 64 dimensional LFBE features are extracted, as in \citep{raeesy2018lstm, ashihara2019neural}.
The experiments were carried out with sampling rates set to both 44.1kHz and 16kHz.

\subsection{Training}
\label{section:training}
Six minor variations of the proposed 1D-CNN architecture, described in Section. \ref{section:1d-cnn}, are trained and tested. 
The variations include, changes to the number of convolutional layers, size of the kernels. 
These variations are summarized in Table \ref{table:architectures}.
It should be noted that Fig. \ref{fig:cnn_arch}, is equivalent to `arch 4' in Table \ref{table:architectures}.

\begin{table}[h]
\caption{Various 1D-CNN architectures experimented in the current work. The `Conv. Layers' column contains a list of tuples, each of which signifies a layer, and convey the size of the kernel and the corresponding number of filters respectively.}
\label{table:architectures}
\begin{center}
\vspace{0.25cm}
\begin{tabular}{lll}
\hline
Architecture & Conv. Layers                              & Dense Layers \\ \hline
arch1        & (10, 32), (5, 64)                    & (1024)       \\
arch2        & (20, 32), (10, 64)                   & (1024)       \\
arch3        & (10, 32), (10, 32), (5, 64), (5, 64)   & (1024)       \\
arch4        & (20, 32), (20, 32), (10, 64), (10, 64) & (1024)       \\
arch5        & (10, 32), (10, 32), (5, 64), (5, 64)   & (1024, 512)   \\
arch6        & (20, 32), (20, 32), (10, 64), (10, 64) & (1024, 512)  
\end{tabular}
\end{center}
\end{table}

An LSTM network, is considered the state-of-the-art for the current task. 
It consists of two hidden layers consisting of 64 memory cells each, as detailed in \cite{raeesy2018lstm}.
All these systems are evaluated for both the wTIMIT and CHAINS datasets.
More details can be found in Section. \ref{section:results}.

\subsection{Scoring}
Testing is performed at the utterance level.
The scores are reported in terms of precision (Pre), recall (Re), F1 score (F1) for each class and the overall accuracy (Acc).
Utterance level decision is obtained by comparing the mean of the posteriors at the output of the network, computed across all frames in an utterance, for both classes.
The class with the higher mean posterior is considered the class of the input utterance.

\begin{table}[]
\caption{Evaluation of different 1D-CNN architectures for the proposed QSE feature, under two sampling rates, for the wTIMIT dataset.}
\label{table:evaluation_1}
\begin{center}
\vspace{0.25cm}
\begin{tabular}{llrrrrrrr}
\hline
                                                                 &               & \multicolumn{3}{c}{\textbf{Normal}}                                                                  & \multicolumn{3}{c}{\textbf{Whisper}}                                                                 & \multicolumn{1}{l}{}                  \\ \cline{3-9}
\textbf{\begin{tabular}[c]{@{}l@{}}Sampling\\ Rate\end{tabular}} & \textbf{Arch} & \multicolumn{1}{c}{\textbf{Pre}} & \multicolumn{1}{c}{\textbf{Re}} & \multicolumn{1}{c}{\textbf{F1}} & \multicolumn{1}{c}{\textbf{Pre}} & \multicolumn{1}{c}{\textbf{Re}} & \multicolumn{1}{c}{\textbf{F1}} & \multicolumn{1}{c}{\textbf{Acc (\%)}} \\ \hline
\multirow{6}{*}{44 kHz}                                          & arch1         & 0.9904                           & 0.9771                          & 0.9837                          & 0.9773                           & 0.9905                          & 0.9839                          & 98.38                               \\
                                                                 & arch2         & 0.9402                           & 0.9937                          & 0.9662                          & 0.9933                           & 0.9367                          & 0.9641                          & 96.52                               \\
                                                                 & \textbf{arch3}         & 0.9743                           & 0.9889                          & 0.9816                          & 0.9887                           & 0.9739                          & 0.9813                          & \textbf{98.14}                               \\
                                                                 & arch4         & 0.8882                           & 0.9984                          & 0.9401                          & 0.9982                           & 0.8741                          & 0.9320                          & 93.63                               \\
                                                                 & arch5         & 0.9734                           & 0.9850                          & 0.9792                          & 0.9848                           & 0.9731                          & 0.9789                          & 97.90                               \\
                                                                 & arch6         & 0.9327                           & 0.9976                          & 0.9641                          & 0.9974                           & 0.9279                          & 0.9614                          & 96.28                               \\ \hline
\multirow{6}{*}{16 kHz}                                          & arch1         & 0.9546                           & 0.9984                          & 0.9760                          & 0.9983                           & 0.9525                          & 0.9749                          & 97.55                               \\
                                                                 & arch2         & 0.9547                           & 1.0000                          & 0.9768                          & 1.0000                           & 0.9525                          & 0.9757                          & 97.63                               \\
                                                                 & arch3         & 0.9598                           & 1.0000                          & 0.9795                          & 1.0000                           & 0.9580                          & 0.9786                          & 97.90                               \\
                                                                 & \textbf{arch4}         & 0.9891                           & 0.9972                          & 0.9931                          & 0.9972                           & 0.9889                          & 0.9931                          & \textbf{99.31}                               \\
                                                                 & arch5         & 0.8883                           & 1.0000                          & 0.9409                          & 1.0000                           & 0.8741                          & 0.9328                          & 93.71                               \\
                                                                 & arch6         & 0.7794                           & 1.0000                          & 0.8760                          & 1.0000                           & 0.7165                          & 0.8349                          & 85.84 \\ \hline                             
\end{tabular}
\end{center}
\end{table}

\section{Results and Discussion}
\label{section:results}

In the following sections we discuss the, 
\begin{itemize}
    \item evaluation of the QSE feature and the architectures listed in Table \ref{table:architectures},
    \item comparison of the QSE and MFCC features,
    \item evaluation of QSE derived from other quarters of the spectrum,
    \item comparison of the proposed system with the state-of-the-art,
    \item evaluation of the QSE feature in the presence of white noise.
\end{itemize}

\subsection{Evaluating the QSE-1D-CNN System}
In this section we discuss the influence of varying the kernel size, number of convolutional layers, number of dense layers of the 1D-CNN. 
We also vary the sampling rate, when computing the QSE feature and examine it's influence in the performance.
Both the wTIMIT and the CHAINS dataset are used for this evaluation, but only the results corresponding to the wTIMIT dataset are reported in Table \ref{table:evaluation_1}. 
This is because an accuracy of 100\% was achieved for the CHAINS dataset across all architectures and sampling rates. 
Hence, this (redundant) information is not included in the table.
The various observations and conclusions from the results follow. 

\subsubsection{Kernel Size}
The influence of the kernel size can be seen when comparing, arch1, arch3 and arch5 with arch2, arch4 and arch6 respectively. In most cases, the accuracy decreases when the kernel size is large. This can be attributed to the fact that the long kernels will smoothen out the spectral envelope of normal speech such that the filter-map resembles that of whispered speech. At the same time, it should be mentioned that, if the size is too small, it will not pick the shape of the harmonics accurately. This is reflected in the results.

\begin{figure}[h]%
\centering
\includegraphics[width=\textwidth]{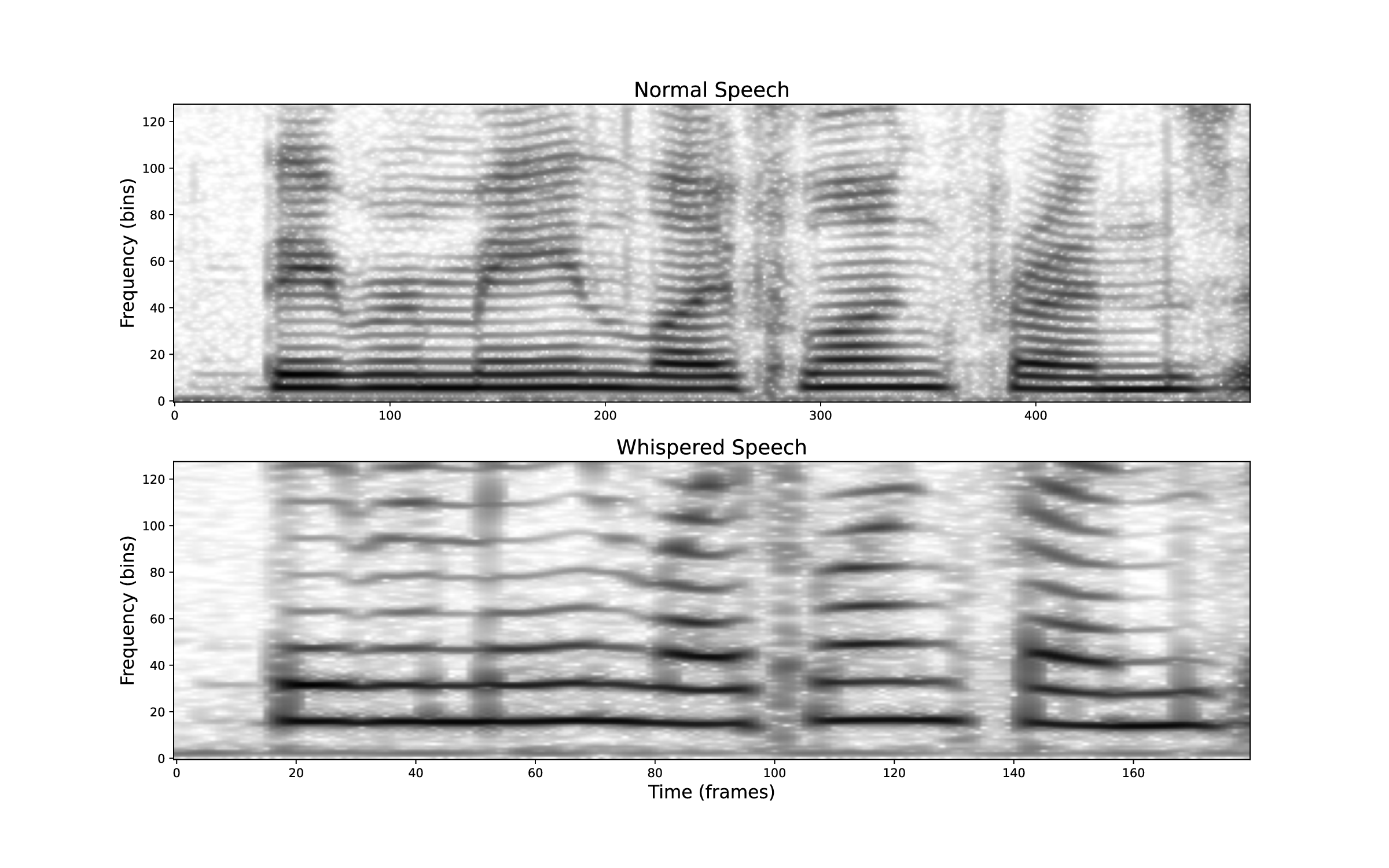}
\caption{A section of the spectrogram computed with 1024 point FFT, spanning from bins 1 to 128, corresponding to normal speech, sampled at 44.1kHz (above) and 16kHz (below).}
\label{fig:reduced_samplingrate}
\end{figure}

\subsubsection{Dense Layers}
Having more than one dense layer could aid in learning the feature maps better. 
But it also means that more data maybe required. 
It can be seen from the table that arch5 and arch6, do not have better scores.
This also means that the additional layers would only negatively contribute to the computational complexity of the system. 

\subsubsection{Sampling Rate}
When the sampling rate is reduced to 16kHz, keeping the size of the FFT constant, we obtain the best accuracy of 99.31 (arch4)\%. 
It can be explained as follows.
For normal speech, when the sampling rate is set to 44.1kHz, and when the first 128 bins are retained, the input to the 1D-CNN is in essence a low pass filtered spectrum with frequencies spanning from 0 to $\sim$5.5kHz.
When 128 bins are retained on a 16kHz signal, the cut-off frequency of the low pass filter is 2kHz. 
The major difference is that, in the latter case, only a few pitch harmonics would be present (up to 2kHz). 
This tells us that a few prominent pitch harmonics are better than including more harmonics that do not offer any additional information. 
Figure \ref{fig:reduced_samplingrate} illustrates the difference in the spectrogram (or spectral envelope) when the sampling rate is reduced for normal speech. 

Another possible explanation pertains to the first formant (F1). 
It is established that amplitude of F1 is much lower for whispered speech \citep{wilson1985comparative, wenndt2002study}. 
It is also established that the F1 occurs at a higher frequency, with increased bandwidth, in whispered speech \citep{jovivcic1998formant}. 
The quartered spectrum of the down-sampled signal is dominated by the first formant.
Hence, the 1D-CNN could be forming yet another distinguishing feature map by learning the prominent shape of the first formant, along with the pitch harmonics.

To ensure the validity of the QSE, we tried this same (best) setup, but with the halved spectral envelope (256 bins), rather than quartered (128 bins).
It returned an accuracy of 98.90\%, which is lower than what is obtained with only the quartered spectral envelope (99.31\%). 
Additionally, using a 256 dimensional input increases the computational complexity.
Henceforth, across all experiments, the features are extracted with the sampling rate set to 16kHz.

\begin{figure}[h]%
\centering
\includegraphics[width=\textwidth]{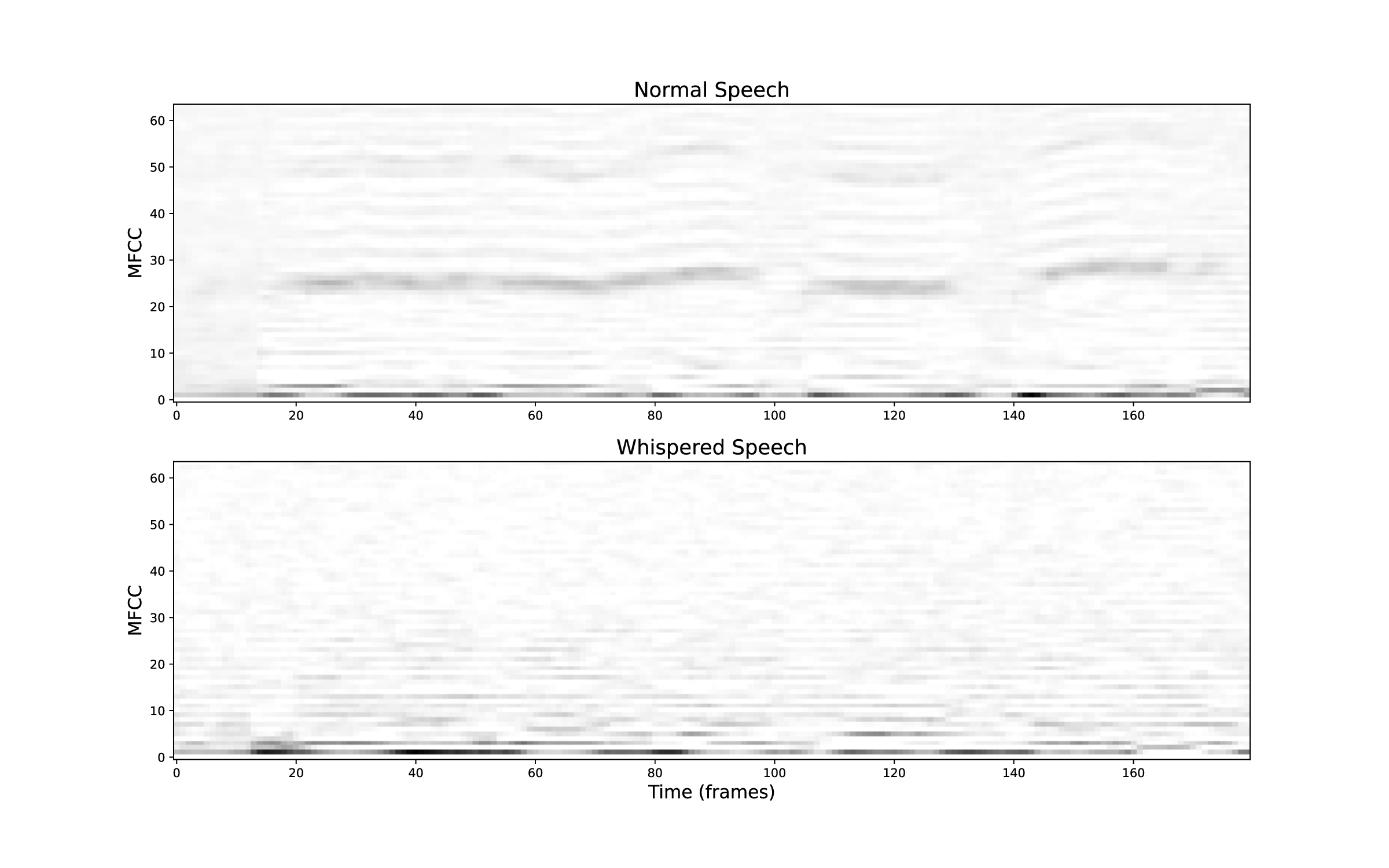}
\caption{MFCC features extracted from normal and whispered speech, sampled at 16kHz.}
\label{fig:mfcc_full}
\end{figure}

\begin{figure}[h]%
\centering
\includegraphics[width=\textwidth]{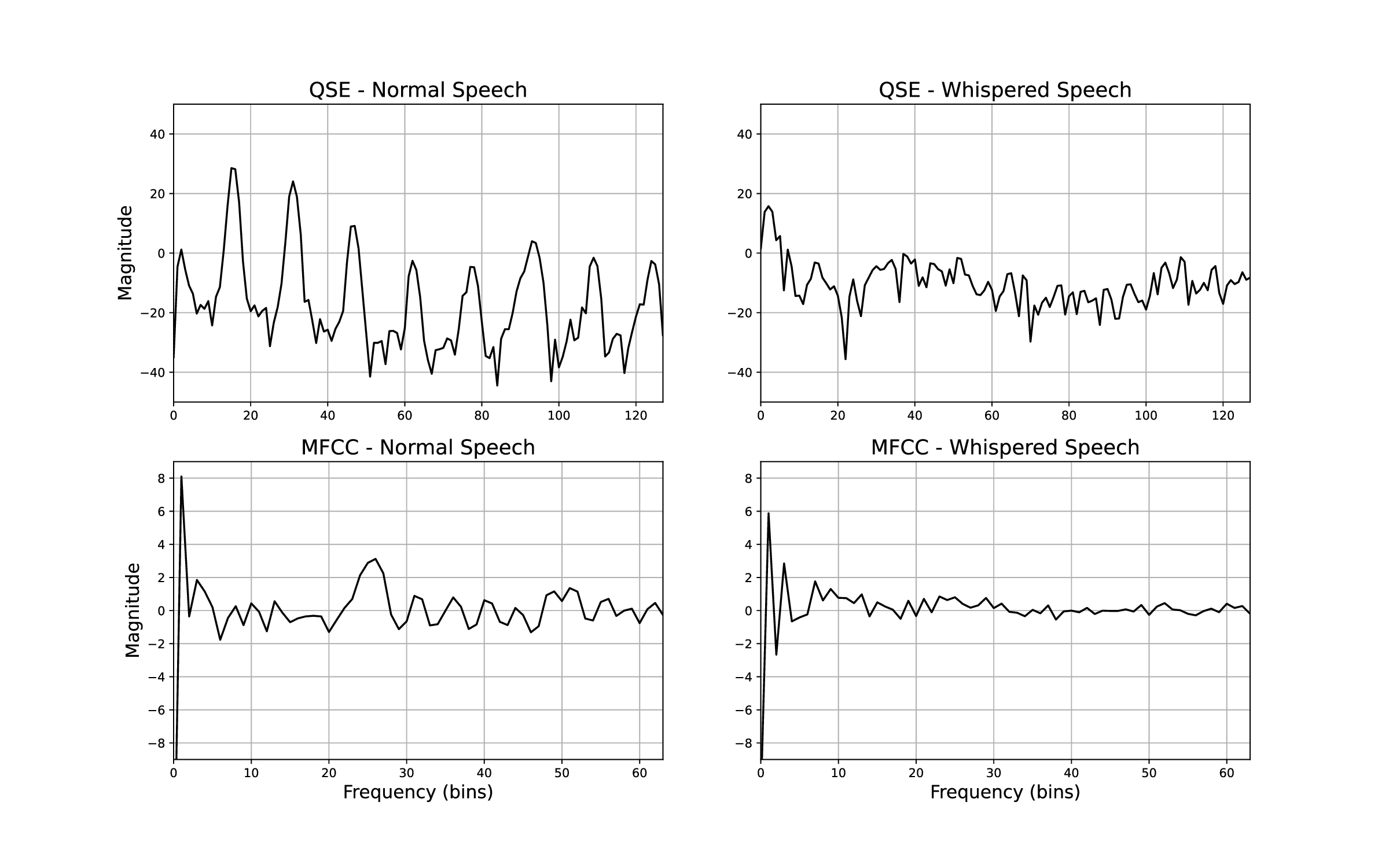}
\caption{Difference between QSE and MFCC feature, shown for both normal and whispered speech, sampled at 16kHz.}
\label{fig:qse-mfcc}
\end{figure}

\begin{table}[]
\caption{Evaluation of different 1D-CNN architectures using MFCC as the feature, for the wTIMIT dataset sampled at 16kHz.}
\label{table:evaluation_2}
\begin{center}
\vspace{0.25cm}
\begin{tabular}{lrrrrrrr}
\hline
              & \multicolumn{3}{c}{\textbf{Normal}}                                                                  & \multicolumn{3}{c}{\textbf{Whisper}}                                                                 & \multicolumn{1}{l}{}             \\ \cline{2-8}
\textbf{Arch} & \multicolumn{1}{c}{\textbf{Pre}} & \multicolumn{1}{c}{\textbf{Re}} & \multicolumn{1}{c}{\textbf{F1}} & \multicolumn{1}{c}{\textbf{Pre}} & \multicolumn{1}{c}{\textbf{Re}} & \multicolumn{1}{c}{\textbf{F1}} & \multicolumn{1}{c}{\textbf{Acc (\%)}} \\ \hline
arch1         & 0.7931                           & 0.9968                          & 0.8834                          & 0.9957                           & 0.7395                          & 0.8487                          & 86.83                          \\
arch2         & 0.8008                           & 0.9976                          & 0.8884                          & 0.9968                           & 0.7514                          & 0.8569                          & 87.46                          \\
arch3         & 0.8847                           & 0.9945                          & 0.9364                          & 0.9937                           & 0.8702                          & 0.9278                          & \textbf{93.24}                          \\
arch4         & 0.8538                           & 0.9968                          & 0.9198                          & 0.9962                           & 0.8290                          & 0.9049                          & 91.30                          \\
arch5         & 0.7689                           & 0.9968                          & 0.8682                          & 0.9955                           & 0.6999                          & 0.8219                          & 84.85                          \\
arch6         & 0.8395                           & 0.9968                          & 0.9115                          & 0.9961                           & 0.8092                          & 0.8930                          & 90.31     \\ \hline                    
\end{tabular}
\end{center}
\end{table}

\subsection{Comparing QSE and MFCC}
In order to explicitly compare the impact of the QSE feature in the proposed QSE-1D-CNN system, we also employ 1D-CNN to learn the envelope of MFCC features.
We extract 64 dimensional MFCC, without velocity or acceleration features \cite{ashihara2019neural}. 
We offer results for all the architectures listed in Table \ref{table:architectures}, in Table \ref{table:evaluation_2}.
It can be seen that out of the 6 experimental setups, only two offer improvement over the QSE feature.
On an average, there is a negative difference of 6.33\% in accuracy, when using the MFCC instead of QSE, across all architectures.
The difference in performance can be attributed to the fact that the pitch harmonics are not that prominent in the envelope of the MFCC (across frequency, in one time frame).
This is illustrated in Fig. \ref{fig:mfcc_full}.
The Mel filters remove some details with respect to the pitch harmonics.
This difference can be seen when comparing Fig. \ref{fig:mfcc_full} with Fig. \ref{fig:spectrogram_full} and \ref{fig:spectrogram_section}.
In Fig. \ref{fig:qse-mfcc}, we show the differences between the QSE and MFCC feature for normal and whispered speech. 
It can be seen that the QSE packs much more visible information about the pitch harmonics than the MFCC. 
Yet, the MFCC does have some discriminating information, which is captured by the 1D-CNN and reflected in the scores.

\subsection{Evaluating QSE Extracted from Other Quarters of the Spectrum}
When considering the use of the first quarter of the spectrum, a question naturally arises as to how the other quarters perform under the same experimental setups, and whether it is the first quarter that offers the best results.
This hypothesis is evaluated in this section. 
To evaluate this, we take the best architecture from the previous experiments, namely `arch4', and vary the computation of the QSE feature by selecting different quarters of the spectral envelope.
The results are summarized in Table \ref{table:evaluation_quarters}.

From the results, it can be seen that there is a drop in performance when anything other than the first quarter is used. 
This is the case in both datasets --- wTIMIT and CHAINS --- although the difference is more evident in the wTIMIT dataset. 
Another trend in the results must also be observed. 
The performance decays as we move away from the first quarter.
The second quarter offers better performance than the third or fourth, but still lesser than the first.
When the third or fourth quarter is used as a feature for the wTIMIT dataset, the confusions are evident from the results. 
The results are biased either towards whispered speech (third quarter), or normal speech (fourth quarter).
This can be attributed to the fact that as we move away from the first quarter, the pitch harmonics are less evident in the spectral envelope.
This can be observed in Fig. \ref{fig:spectrogram_full} as well.

\begin{table}[]
\caption{Evaluating QSE extracted from all the quarters of the spectrum, for the CHAINS and wTIMIT dataset}
\label{table:evaluation_quarters}
\begin{center}
\vspace{0.25cm}
\begin{tabular}{llrrrrrrr}
\hline

                        & \textbf{}        & \multicolumn{3}{c}{\textbf{Normal}}                                                                  & \multicolumn{3}{c}{\textbf{Whisper}}                                                                 & \multicolumn{1}{l}{}             \\ \cline{3-9}
\textbf{Dataset}        & \textbf{Quarter} & \multicolumn{1}{c}{\textbf{Pre}} & \multicolumn{1}{c}{\textbf{Re}} & \multicolumn{1}{c}{\textbf{F1}} & \multicolumn{1}{c}{\textbf{Pre}} & \multicolumn{1}{c}{\textbf{Re}} & \multicolumn{1}{c}{\textbf{F1}} & \multicolumn{1}{c}{\textbf{Acc (\%)}} \\ \hline
\multirow{4}{*}{CHAINS} & Q1               & 1.0000                           & 1.0000                          & 1.0000                          & 1.0000                           & 1.0000                          & 1.0000                          & \textbf{100.00}                         \\
                        & Q2               & 0.9729                           & 0.9875                          & 0.9801                          & 0.9873                           & 0.9725                          & 0.9798                          & 98.00                          \\
                        & Q3               & 0.8077                           & 0.9975                          & 0.8926                          & 0.9967                           & 0.7625                          & 0.8640                          & 88.00                          \\
                        & Q4               & 0.8230                           & 0.9650                          & 0.8884                          & 0.9577                           & 0.7925                          & 0.8673                          & 87.88                          \\ \hline
\multirow{4}{*}{wTIMIT} & Q1               & 0.9891                           & 0.9972                          & 0.9931                          & 0.9972                           & 0.9889                          & 0.9931                          & \textbf{99.31}                          \\
                        & Q2               & 0.9354                           & 0.9162                          & 0.9257                          & 0.9178                           & 0.9367                          & 0.9271                          & 92.64                          \\
                        & Q3               & 0.0000                           & 0.0000                          & 0.0000                          & 0.4996                           & 1.0000                          & 0.6663                          & 49.96                          \\
                        & Q4               & 0.5210                           & 1.0000                          & 0.6851                          & 1.0000                           & 0.0792                          & 0.1467                          & 54.00      \\ \hline                   
\end{tabular}
\end{center}
\end{table}

\subsection{Evaluation of the Proposed and the State-of-the-art}

The state-of-the-art for the current task uses, log-filterbank energies (LFBE) as a feature, which are trained on a long short-term memory (LSTM) network \cite{raeesy2018lstm, ashihara2019neural}.
We compare this system with the proposed QSE-1DCNN system. 
We also train the LSTM with MFCC and the QSE features, to compare the features specifically. 
We do this for both datasets. 
The results are provided in Table \ref{table:lstm}. 
Yet again, perfect accuracy was obtained for the CHAINS  dataset hence, the results are not provided in the table.

From Tables \ref{table:evaluation_1} and \ref{table:lstm}, we can see that the proposed QSE-1DCNN system offers a better performance when compared the LFBE-LSTM system.
Amongst the three features trained using the LSTM system, we again note that the QSE offers better performance than LFBE and MFCC.
Where, the LFBE offers better performance than the MFCC.
It should also be mentioned that while the LSTM network offers near equivalent classification performance, it is more computationally intensive than the proposed 1D-CNN network.

\begin{table}[]
\caption{Evaluation of the state-of-the-art LSTM network for the three features --- LFBE, MFCC and QSE --- for the wTIMIT dataset.}
\label{table:lstm}
\begin{center}
\vspace{0.25cm}
\begin{tabular}{lrrrrrrr}
\hline
                 & \multicolumn{3}{c}{\textbf{Normal}}                                                                  & \multicolumn{3}{c}{\textbf{Whisper}}                                                                 & \multicolumn{1}{l}{}                  \\ \cline{2-8}
\textbf{Feature} & \multicolumn{1}{c}{\textbf{Pre}} & \multicolumn{1}{c}{\textbf{Re}} & \multicolumn{1}{c}{\textbf{F1}} & \multicolumn{1}{c}{\textbf{Pre}} & \multicolumn{1}{c}{\textbf{Re}} & \multicolumn{1}{c}{\textbf{F1}} & \multicolumn{1}{c}{\textbf{Acc (\%)}} \\ \hline
LFBE             & 0.9782                           & 0.9903                          & 0.9842                          & 0.9902                           & 0.9779                          & 0.9840                          & 98.41                               \\
MFCC             & 0.9426                           & 0.9972                          & 0.9692                          & 0.9971                           & 0.9391                          & 0.9672                          & 96.82                               \\
QSE              & 0.9863                           & 0.9945                          & 0.9904                          & 0.9944                           & 0.9862                          & 0.9903                          & \textbf{99.03}       \\ \hline                       
\end{tabular}
\end{center}
\end{table}

\subsection{Evaluation under Noise}
In this section, we evaluate the performance of the proposed QSE-1D-CNN system in the presence of white noise under various signal to noise ratios (SNR).
We add white noise to both the train and test data with a specific SNR.
The results are presented in Table \ref{table:noise}.
It can be seen from the table that even when there is considerable noise in the data, the proposed system performs particularly well with accuracies above 95\%.
The CHAINS dataset achieves perfect accuracy and hence, is not added in the table. 

\begin{table}[]
\caption{Evaluation under white noise at 0dB, 5dB and 10dB SNR, for the wTIMIT dataset.}
\label{table:noise}
\begin{center}
\vspace{0.25cm}
\begin{tabular}{rrrrrrrr}
\hline
\multicolumn{1}{l}{}             & \multicolumn{3}{c}{\textbf{Normal}}                                                                  & \multicolumn{3}{c}{\textbf{Whisper}}                                                                 & \multicolumn{1}{l}{}                  \\ \cline{2-8}
\multicolumn{1}{l}{\textbf{SNR (dB)}} & \multicolumn{1}{c}{\textbf{Pre}} & \multicolumn{1}{c}{\textbf{Re}} & \multicolumn{1}{c}{\textbf{F1}} & \multicolumn{1}{c}{\textbf{Pre}} & \multicolumn{1}{c}{\textbf{Re}} & \multicolumn{1}{c}{\textbf{F1}} & \multicolumn{1}{c}{\textbf{Acc (\%)}} \\ \hline
0                                & 0.9539                           & 0.9421                          & 0.9480                          & 0.9426                           & 0.9544                          & 0.9485                          & 94.82                               \\
5                                & 0.9062                           & 1.0000                          & 0.9508                          & 1.0000                           & 0.8963                          & 0.9453                          & 94.82                               \\
10                               & 0.9602                           & 0.9972                          & 0.9783                          & 0.9971                           & 0.9585                          & 0.9774                          & 97.79      \\ \hline                        
\end{tabular}
\end{center}
\end{table}

\section{Conclusion}
An efficient algorithm based on quartered spectral envelope and 1D-CNN, that captures the inherent characteristic differences between whispered and normal speech is proposed.
These inherent characteristics are the presence of pitch harmonics and the first formant present in normal speech.
They are evident from the Fourier spectrum and spectrogram.
The validity of using the first quarter, instead of the other three, is proven by experimental means.
Consistent results are obtained when the proposed system is evaluated with two datasets --- wTIMIT and CHAINS.
It performs on par with the state of the art system based on LFBE-LSTM.
The proposed QSE feature performs better than MFCC across all experiments.
Since the system is based on 1D-CNN, rather than 2D-CNN, it requires less data and less computational power.
It's robustness is also verified, by evaluation under white noise.
The current work can be extended to, (i) finding the performance of the algorithm in the presence of more complex noise, and (ii) real time classification using the proposed feature and architecture.

\section*{Declarations}
\textbf{Funding:} Not Applicable. This research was not funded.

\vspace{0.2cm}
\noindent
\textbf{Conflicts of Interest:}
The authors declare that are no conflicts of interest.




\vspace{0.2cm}
\noindent
\textbf{Availability of data and material:}
The data used in the current work, the wTIMIT and the CHAINS dataset, are available, on reasonable request, from the authors of \cite{lim2011computational} and \cite{cummins2006chains} respectively.

\vspace{0.2cm}
\noindent
\textbf{Code availability:} Code will be available from the corresponding author, on reasonable request.

\end{document}